\documentclass{article}
\usepackage{epsfig}
\def \cmsq           {\hbox{cm$^{-2}$}}
\def \deg          {\ifmmode ^{\circ}\else $^\circ$\fi}  

\def \flam         {\hbox{ergs s$^{-1}$ cm$^{-2}$ \AA$^{-1}$}}  

\def \kms          {\rm{\hbox{km s$^{-1}$}}}
\def \lam          {$\lambda$}

\def \Msun          {\hbox{M$_{\odot}$}}
\def \pcc           {\hbox{cm$^{-3}$}}

\def \zaz          {{$z_a\kern -1.5pt \approx\kern -1.5pt z_e$}}
\def \zgz          {{$z_a\kern -1.5pt >\kern -1.5pt z_e$}}
\def \zllz         {{$z_a\kern -3pt \ll\kern -3pt z_e$}}

\def \lap{\mathrel{\hbox{\rlap{\hbox{\lower4pt\hbox{$\sim$}}}\hbox{$<$}}}}
\def\gap{\mathrel{\hbox{\rlap{\hbox{\lower4pt\hbox{$\sim$}}}\hbox{$>$}}}}%
\begin{document}
\twocolumn
\section*{Intrinsic AGN Absorption Lines}

Any gaseous material along our line of sight to distant 
quasars or, more generally, active galactic nuclei (AGNs) will 
absorb light according to the amounts and range of ions 
present. Strong absorption lines are common in rest-frame 
UV spectra of AGNs due to a variety of 
resonant transitions, for example 
the HI Lyman series lines (most notably Ly$\alpha$ \lam 1216) 
and high-ionization doublets like CIV \lam\lam 1549,1551.    
The lines are called {\it intrinsic} if the absorbing gas is 
physically related to the AGN, e.g. if the absorber resides  
broadly within the radius of the AGN's surrounding ``host'' galaxy. 
Intrinsic absorption lines are thus valuable probes of the 
kinematics, physical conditions and elemental abundances in 
the gas near AGNs. 

Studies of intrinsic absorbers have historically emphasized 
quasar broad absorption lines (BALs), which 
clearly identify energetic outflows from the central engines. 
Today we recognize a wider variety of intrinsic lines 
in a wider range of objects. For example, we now know that 
Seyfert 1 galaxies (the less luminous cousins of quasars) have 
intrinsic absorption. We also realize that intrinsic lines can 
form in a range of AGN environments --- from the dynamic inner 
regions like the BALs, to the more quiescent outer host galaxies 
$>$10 kpc away. One complicating factor is that most AGNs also 
have absorption lines due to unrelated gas or galaxies far 
from the AGN (see {\sc quasars: absorption lines}). Part of the 
effort, therefore, is to identify the intrinsic lines and 
locate their absorbing regions relative to 
the AGNs. Empirical line classifications are a good 
starting point for this work. 

\subsection*{Empirical Line Types}

AGN absorption lines are usually classified according to the 
widths of their profiles. These classes 
separate the clearly intrinsic broad lines from the many 
others of uncertain origin. The main empirical classes are 
1) the BALs, 2) the narrow absorption lines (NALs), and 3) 
the intermediate ``mini-BALs.'' 

Broad absorption lines (Fig. 1) are blueshifted relative to 
the AGN emission lines, implying outflow velocities from near 
0 \kms\ to as much as $\sim$60,000 \kms\ ($\sim$0.2$c$). A 
representative line width is $\sim$10,000 \kms , although 
there is considerable diversity among BAL profiles. 
Velocity widths $\gap$3000 \kms\ and blueshifted 
velocity extrema $\gap$5000 \kms\ are usually considered 
minimum requirements for classification as a BAL. 
Some BALs have several distinct absorption troughs, while others 
are strictly ``detached'' from the emission lines --- such that 
the absorption appears only at blueshifts exceeding 
several thousand \kms . 

Narrow absorption lines (Fig. 2) have 
widths less than a few hundred \kms . 
NALs with absorption redshifts, $z_a$, within $\pm$5000 \kms\ 
of the emission redshift, $z_e$, are called ``associated'' (or 
\zaz ) lines because of their likely physical connection to the 
AGN. In general, NALs can appear at 
redshifts from \zaz\ to $z_a\approx 0$. Many of these systems 
are not related to the AGNs.

\begin{center}
\vskip -1.3cm 
\hskip -0.76cm
\epsfig{file=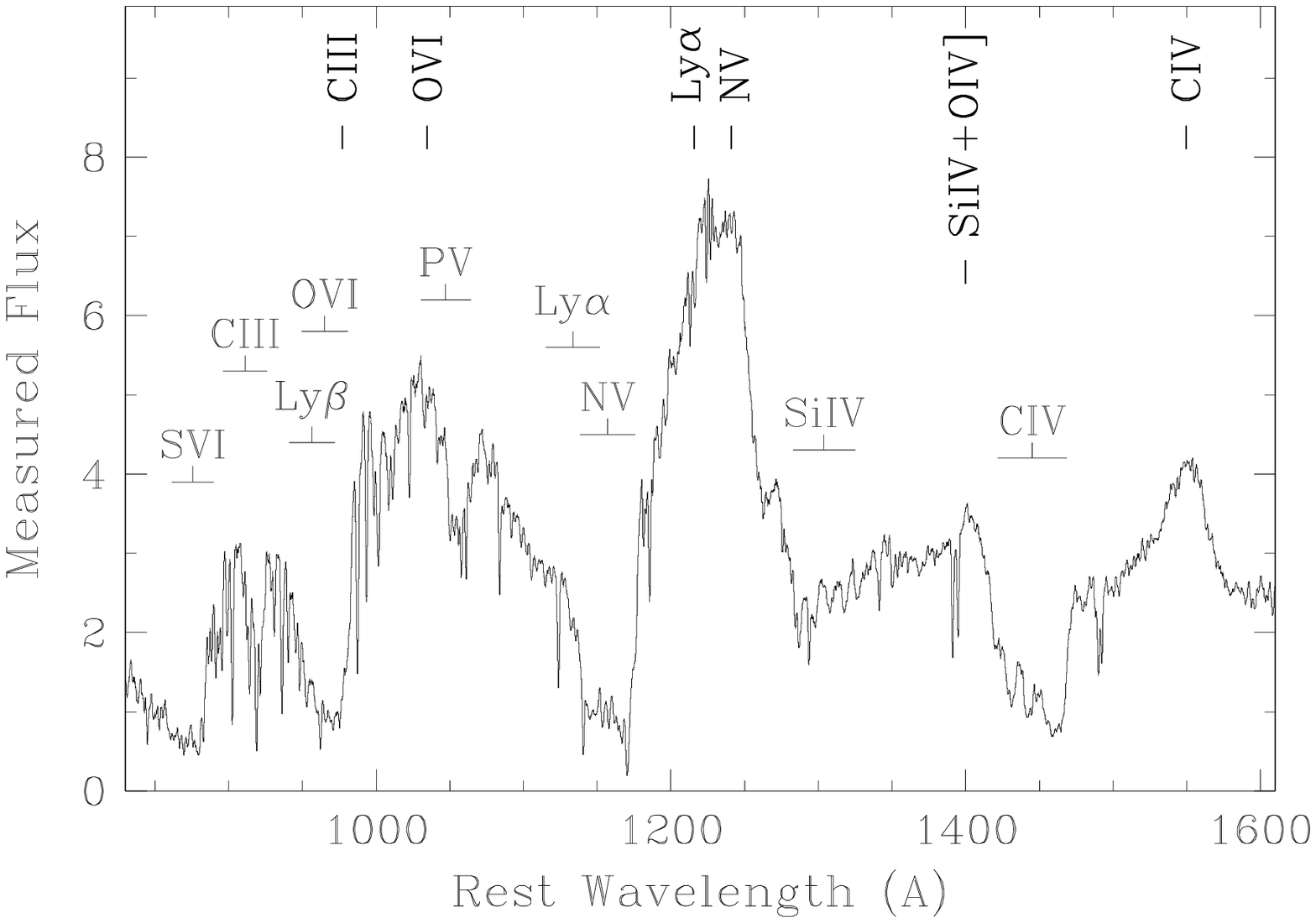,width=9.2cm}
\end{center}
\vskip -5.8cm
\noindent Figure 1. --- Detached BALs in the quasar 
PG 1254+047. The BALs are labeled just above the 
spectrum. The wavelengths of prominent broad emission lines are 
marked across the top. The Flux has units 10$^{-15}$ \flam .
\medskip

\begin{center}
\vskip -1.3cm 
\hskip -0.76cm
\epsfig{file=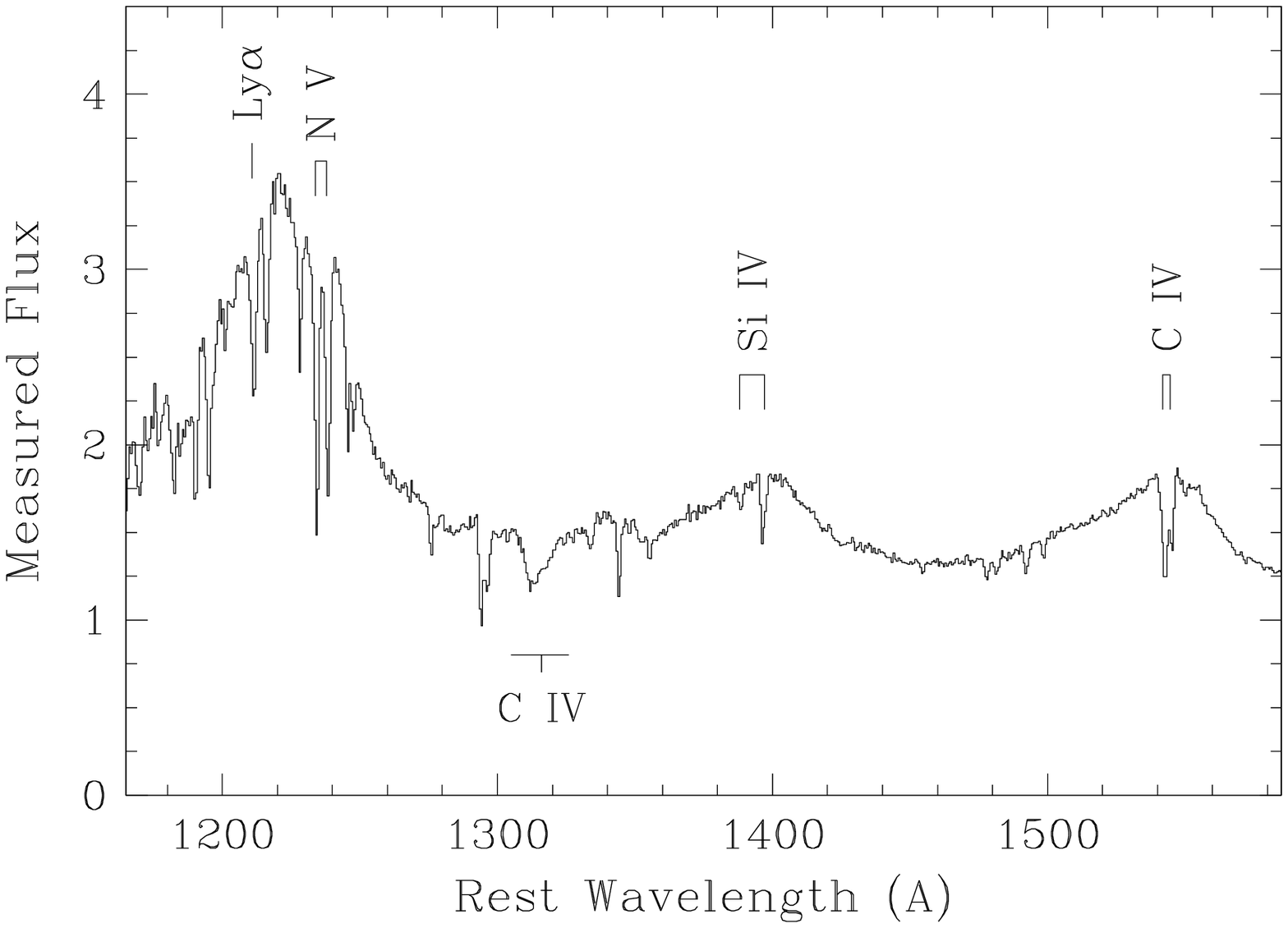,width=9.2cm}
\end{center}
\vskip -5.8cm
\noindent Figure 2. --- Intrinsic absorption in the quasar PG 0935+417. 
A system of associated NALs is labeled above the spectrum, with open brackets 
showing the doublet separations. A mini-BAL due to CIV blueshifted by 
$\sim$51,000 \kms\ is labeled below. The Flux has units 10$^{-15}$ \flam .
\medskip

Absorption lines with intermediate widths between the BALs 
and NALs are increasingly called mini-BALs (Fig. 2). 
Their strictly blueshifted profiles appear smooth 
and BAL-like in high resolution spectra, and 
their centroid velocities span the same range as the BAL troughs 
(from near 0 \kms\ to almost 0.2$c$). Mini-BALs evidently 
form in outflows similar to the BALs.  

\subsection*{Identifying Intrinsic NALs}

The first evidence that some NALs are intrinsic came from 
statistical tendencies, namely, 1) quasar NALs appear with 
greater frequency near the emission-line redshift, and 2) the
strengths of these \zaz\ systems correlate with the quasars' radio 
properties. The first tendency might be explained by external  
galaxies clustering around quasars, but the second 
clearly demonstrates a physical relationship 
to the quasars themselves. 

\begin{center}
\vskip -0.6cm 
\hskip -0.76cm
\epsfig{file=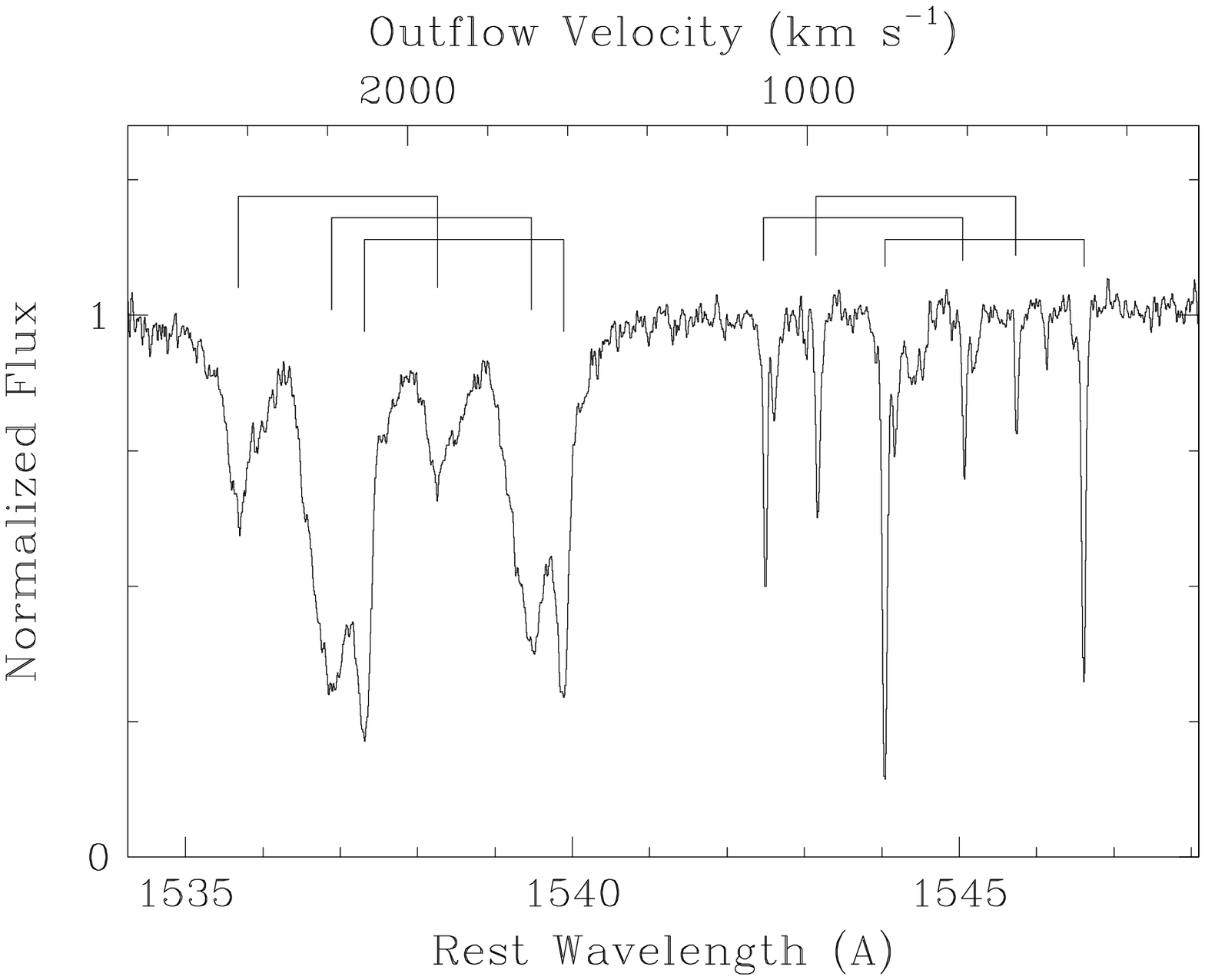,width=9.2cm}
\end{center}
\vskip -5.8cm
\noindent Figure 3. --- High resolution blow-up of the CIV 
NALs in PG 0935+417, revealing multiple components (cf. Fig. 2). 
The open brackets mark the strongest doublet pairs. The outflow 
velocities are appropriate for the short-wavelength doublet members. 
The absorption complex at 1535--1540 \AA\ is intrinsic based on the 
broad line profiles and doublet ratios that imply partial 
line-of-sight coverage.
\medskip

Direct evidence for the intrinsic origin of specific NALs 
has come from spectroscopic indicators, such as 
1) time-variable line strengths, 2) well-resolved line profiles that 
are smooth and broad compared to the thermal velocity, 3) multiplet 
ratios that imply partial coverage of the background light source(s), 
and 4) high space densities ($\gap$100 \pcc ) 
inferred from the presence of excited-state absorption lines. 
These properties signal intrinsic absorption because 
they are most easily understood in terms of the dense and 
dynamic environments near AGNs. Unrelated 
intervening absorbers --- typically inter-galactic 
gas clouds or extended galactic halos --- should generally be 
larger, more quiescent, and less ionized for a given gas 
density. The link between the first 3 properties 
and the near-quasar environment is further strengthened 
by the fact that they are common in BALs and mini-BALs. 
NALs with these characteristics probably also form 
in outflows; they have been measured 
at blueshifted (ejection) velocities from $\sim$0 to 
$\sim$24,000 \kms\ (e.g. Fig. 3).   

\subsection*{Global Covering Factors and the Ubiquity of 
Intrinsic Absorbers}

All varieties of quasars and Seyfert 1 galaxies show some 
type of intrinsic absorption, but different objects 
favor different types of absorbers. 
For example, BALs and other high-velocity ($\gap$3000 \kms ) 
intrinsic lines appear only in quasars and never 
in the Seyfert galaxies. BALs also tend to avoid quasars with 
strong radio emission, while associated NALs seem to favor 
them. Preferences like these are probably tied to the 
unique physical conditions, whereby 
different AGNs drive different types of outflows. 

Intrinsic lines also do not appear in every individual 
spectrum. For example, BALs are detected in only $\sim$12\% 
of quasars. 
The detection rate of associated (and so probably intrinsic) 
NALs in quasars is not well known, but it appears to 
be roughly similar to the BALs. 
Mini-BALs may be somewhat rarer than both BALs and  
associated NALs in quasars, but they can appear in either 
``radio-loud'' or ``radio-quiet'' objects. Mini-BALs 
and intrinsic NALs are both common in Seyfert 1 
galaxies, with at least one of these features appearing 
in 50--70\% of the low-luminosity sources. 

These detection rates (within object classes) depend on viewing angle 
effects and on the global covering factors of the absorbing gas. 
If the objects are randomly oriented, the absorption-line detection 
frequencies should approximately equal the average value of the global 
covering factor, $Q\equiv\Omega /4\pi$, where $\Omega$ is the 
solid angle subtended by the absorber as seen from the central 
light source. In practice, the situation can be more complicated. 
For example, attenuation by BAL gas might bias quasar 
samples against the inclusion of sources with BALs. Thus 
the true global covering factor of BAL regions could be 20--30\%,  
instead of the $\sim$12\% implied by the detection rate. 
Nonetheless, the overall conclusion is that 
intrinsic absorbers are present in all AGNs, with 
global covering factors {\it similar} to the line  
detection frequencies in each type of object. 

\subsection*{Absorber Geometry}

A consistent picture of the geometry has emerged 
in which the absorbing gas resides primarily near 
the equatorial plane of the AGN accretion disk (Fig. 4). 
In particular, spectropolarimetric observations show that 
the continuum light from BAL quasars is more 
polarized than non-BAL quasars. Also, the percent polarization 
is typically much greater inside BAL troughs compared to 
the adjacent continuum. These results are understood in terms 
of light scattering in the disk geometry. Quasars 
viewed close to edge-on exhibit BALs because our sight-line 
intersects the the absorbing gas. These objects are also 
more polarized because the direct (unpolarized) light through 
the BAL region is slightly attenuated; thus the scattered (polarized) 
light makes a relatively larger contribution to their 
measured flux. BAL troughs have higher percent polarizations 
because the direct (unpolarized) light in the troughs is 
more attenuated than the continuum.
Quasars without BALs, viewed from above or below 
the disk plane, have low polarizations because their direct 
(unpolarized) light dominates the measured flux. 

\begin{center}
\vskip 3.5cm \hskip -0.5cm
\epsfig{file=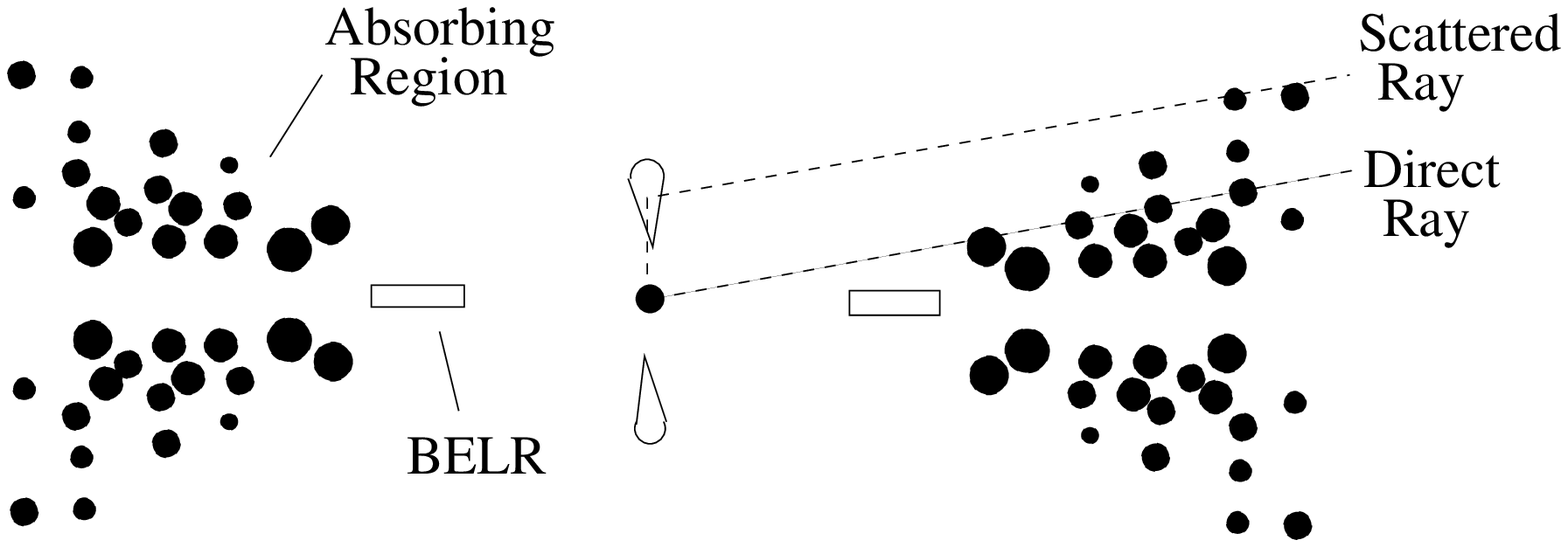,width=8.7cm}
\end{center}
\vskip -4.5cm
\noindent Figure 4. --- Schematic absorber geometry. The dotted 
lines are representative light rays from the continuum source 
(central dot). The open rectangles mark the broad emission 
line region (BELR) near the accretion disk 
plane. The teardrop lobes above and below are putative 
scattering regions near the disk axis. Extended radio jets, 
when present, would lie along this axis. 
\medskip

This picture is also supported by observations showing 
that quasars with reddened spectra, presumably caused by dust 
near the disk plane, are more likely to have BALs and high 
polarizations. Similarly, studies of 
radio-loud quasars show that associated NALs are stronger and 
more common in sources with their radio jets aligned 
perpendicular to our line of sight (such that their unresolved 
inner disks are viewed nearly edge-on). 

\subsection*{Basic Physical Properties}

\subsubsection*{Kinematics}

One surprising property of intrinsic absorbers 
is that none of them have changed velocity between 
observations that now span 10--20 yrs. In one well-studied case, 
distinct features in a quasar BAL show $<$30 \kms\ of movement 
over 5 yrs in the quasar rest frame, implying an 
acceleration of $<$0.02 cm s$^{-2}$. The outflow speeds of 
10,000--20,000 \kms\ are therefore stable to $\lap$0.2\% 
on this time scale.

Another general property is the kinematic complexity (e.g. Figs. 1--3). 
Many intrinsic absorbers have multiple distinct absorption troughs. 
Sometimes BALs, mini-BALs and/or NALs appear together in the same 
spectrum, either blended together or at different 
(non-overlapping) outflow velocities. The narrower lines, i.e. the 
NALs and mini-BALs, generally have small velocity dispersions 
compared to the outflow speeds. Evidently, the outflows 
producing these lines are not continuously accelerated from 
rest along our line of sight. These 
properties all present challenges for the physical models 
discussed below.

\subsubsection*{Column Densities and Partial Coverage}

Estimates of the column densities in each absorbing ion 
are complicated by the fact that many (most?) intrinsic 
absorbers only partially cover 
the background light source(s) along our line(s) of sight. 
Partial coverage might occur if the absorbing regions 
are porous or they are smaller than the emission sources 
in overall projected area (see Fig. 4). 
In any case, partial coverage 
leads to unabsorbed flux filling in 
the bottoms of the measured troughs. The observed 
line intensities thus depend on both the 
line-of-sight coverage fraction, $C_f$, and the 
line optical depths as, 
\begin{equation}
I_v\ =\ (1-C_f)\,I_o\, +\, C_f\,I_o\,e^{-\tau_{v}}
\end{equation}
where $0\leq C_f\leq 1$, $I_v$ and $I_o$ are the observed and 
emitted (unabsorbed) intensities, and $\tau_v$ is the line 
optical depth, at each line velocity $v$. The first term on the 
right side represents the unabsorbed light that fills in the 
troughs. In the limit $\tau_{v}\gg 1$ we have simply, 
\begin{displaymath}
C_f\ =\ 1 - {{I_{v}}\over{I_o}}
\end{displaymath}
Outside of that limit, 
we can compare lines whose true optical depth ratios are fixed by 
atomic constants, such as the HI Lyman lines or doublets like 
CIV~\lam\lam 1548,1550, to 
determine uniquely both the coverage fractions and the true 
optical depths across the line profiles. For the commonly measured 
doublets (like CIV), where the true optical depth ratios are $\sim$2, 
the coverage fraction at each velocity follows from Eqn. 1 such that, 
\begin{displaymath}
C_f\ =\ {{I_1^2 - 2I_1 + 1}\over{I_2 - 2I_1 +1}} 
\end{displaymath}
where $I_1$ and $I_2$ are the observed 
line intensities, normalized by $I_o$, at the same velocity in 
the weaker and stronger line troughs, respectively. 
The column densities, $N$, follow from the line optical depths via,
\begin{displaymath}
N\ =\ {{m_ec}\over{\pi e^2 f\lambda_o}}\,\int\tau_v\ dv
\end{displaymath}
where $f$ and $\lambda_o$ are the line's oscillator strength and
laboratory wavelength. The integral is 
performed over all or part of the line profile. 

This analysis has been applied to well-resolved multiplets 
in NALs and mini-BALs. The derived coverage fractions range 
from $\sim$10\% to 100\%. The  
column densities in commonly measured ions 
(e.g. HI, CIV, NV, SiIV, etc.) are usually in the range 
$10^{13}\lap N\lap 10^{16}$ \cmsq , consistent with the usual 
absence of Lyman edge absorption at 912 \AA . Most BALs 
are too broad and blended for this analysis. They  
probably have $C_f < 1$ in general, but the exact values are 
rarely known. 
Crude estimates suggest that the range of BAL  
column densities is perhaps an order of magnitude larger 
than the well-measured NALs. 

\subsubsection*{Ionization and Total Column Densities}

Intrinsic absorbers are in general highly ionized. Their 
lowest levels of ionization are usually characterized 
by ions such as CIII, NIII, SiIV or CIV. Lower ionization species 
like SiII, FeII, MgII or CII are rarely present. 
The upper limits of ionization are generally not known 
because higher ions have their resonance lines at shorter 
(and more difficult-to-measure) wavelengths. Existing 
data suggest that ionizations up to at 
least NeVIII are common. 

The strong radiative flux of the AGN provides a natural 
ionization source. Theoretical simulations of photoionized 
plasmas in equilibrium with the AGN radiation field are used to 
match the measured column densities, and thus derive the overall 
ionization state(s) and total column densities (in HI + HII). 
The ionization state is described by an ionization parameter, 
$U$, which is the dimensionless ratio of hydrogen-ionizing photon 
to hydrogen particle densities in the absorber:
\begin{displaymath}
U\ \equiv\ {{1}\over{4\pi R^2\, c\, n_H}}\, 
\int_{\nu_{LL}}^{\infty}{{L_{\nu}}\over{h\nu}}\, d\nu 
\end{displaymath}
where $n_H$ is the absorber's hydrogen density, $R$ is its 
radial distance from the central quasar, $L_{\nu}$ is the quasar 
luminosity distribution and $\nu_{LL}$ is the frequency at the 
HI Lyman limit. For a ``typical'' AGN spectral shape we have,
\begin{equation}
U\ \approx\ 0.3\; n_{10}^{-1}\, R_{0.1}^{-2}\, L_{46}
\end{equation}
where $n_{10}$ is the density in units of 
10$^{10}$ \pcc , $R_{0.1}$ is the radial distance 
in units of 0.1~pc, and $L_{46}$ is the AGN luminosity 
relative to $10^{46}$ ergs s$^{-1}$. Typical $U$ values are 
between $\sim$0.01 and $\sim$1. Absorbers at the same 
velocity, or at different velocities in the same spectrum, 
often have a range of $U$ values implying a range 
of densities or radii in the overall absorbing region. 
Derived total column densities are typically 
$N_{\rm H} \approx 10^{19}$ to $10^{21}$ \cmsq\ for NAL regions and 
$\gap$$10^{22}$ \cmsq\ for the BAL gas.  

\subsubsection*{Correlated X-Ray Absorption}

Recent studies have shown that the 
presence of intrinsic lines in the UV correlates with continuous 
(bound-free) absorption in soft X-rays. Evidently, the UV lines are 
just one aspect of the intrinsic absorber phenomenon. The X-ray 
results are important because they reveal generally higher 
ionizations, $U>1$, and much higher total column densities, by 1--2 
orders of magnitude, compared to the UV lines. A critical issue, 
therefore, is the physical relationship between the UV 
and X-ray absorbers. 

The data clearly show that different absorbers (of either type) 
can have a wide range of physical conditions. 
Some well-measured UV absorbers, in particular, have 
ionizations and total column densities
that should produce only minimal X-ray absorption. 
Moreover, the best ensemble data are often inconsistent with 
a 1-zone medium producing all of the measured features. 
Spatially distinct zones with different ionizations and 
column densities are at least sometimes present. 
It is therefore likely that the UV and X-ray absorbers 
are physically related but not, in general, identically the same. 

\subsubsection*{Space Densities and Radial Distances}

Most constraints on the radial distance come indirectly 
from estimates of the space density and ionization 
parameter (via Eqn. 2). The densities are derived in several 
ways. For example, the absence of broad or blueshifted forbidden 
emission lines, e.g. [OIII] \lam 5007, suggests (for 
BAL regions only) that this emission is suppressed by 
collisional deexcitation at high densities. Some AGNs have 
have absorption lines from excited energy states, implying 
significant densities to support the upper 
level populations. In other cases, time-variable 
absorption lines require minimum densities to allow the gas 
to adjust its ionization structure within the variability time. 
This last method assumes that the variability time 
exceeds the recombination time, 
$t_{recomb} \approx\ (n_e \alpha)^{-1}$, where $n_e$ is the 
electron density and $\alpha$ is the recombination rate 
coefficient. Even if the line variability is not caused by 
changes in the ionization, it turns out that dynamical limits 
on clouds moving across our line of sight lead to similar 
$R$ constraints.  

The overall results imply densities from $\lap$7 \pcc\ to $>$10$^6$ \pcc\ 
and corresponding distances from $\gap$300 kpc to $\lap$10 pc in 
different sources. The largest confirmed distances apply only to 
narrow and non-variable NALs. Lines with clear dynamic 
signatures, e.g. the BALs, mini-BALs and some NALs, probably form 
closer to the AGN than even the smallest of these upper limits. 
The minimum radial distance, $R_{min}$, is set by the fact that 
many intrinsic absorbers, e.g. most BALs, suppress both the 
continuum and broad line emissions. The absorber distances therefore 
cannot be much less than the broad emission line region radius, which 
is known independently to scale with the AGN luminosity, such that, 
\begin{equation}
R_{min}\ \approx\ 0.1\, \left({L_{46}}\right)^{1\over 2}\ \ {\rm pc}
\end{equation}
where $L_{46}\approx 1$ is ``typical'' for quasars. 
(Values of $L_{46}$ can actually range from $\lap$0.001 in 
Seyfert nuclei to $>$100 in the most luminous quasars.) 
If there is absorbing gas near this minimum radius, the densities   
could reach $\sim$$10^{10}$ \pcc\ (see Eqn. 2). 

\subsection*{Physical Models}

Most physical models of intrinsic absorbers feature 
a wind-disk geometry similar to Fig. 4. The outflowing gas 
stays close to the accretion-disk plane, and we observe 
it (via blueshifted absorption lines) only if our sight line(s) 
to the emission source(s) intersect wind material. The disk 
provides a likely source for the wind material, and its  
acceleration to high speeds probably occurs via radiation 
pressure from the central emission source. 
The outward transfer of angular momentum 
in these winds might facilitate the {\it inward} flow of matter 
in the accretion disk, thus promoting the growth and 
fueling of the central black hole. 

The measured line velocities 
suggest that the outflows often originate near the 
minimum radius implied by Eqn. 3. In particular, the radial 
acceleration of a wind driven by radiation pressure from a central 
point source with luminosity $L$ and mass $M$ is, 
\begin{displaymath}
{{v\ dv}\over{dR}} \ = \ {{f_L L}\over{4\pi R^2 c m_{\rm H} N_{\rm H}}}\ - 
{{G M}\over{R^2}}
\end{displaymath}
where $f_L$ is the 
fraction of the luminosity absorbed or scattered in the wind. 
Integrating this equation from $R$ to infinity yields the terminal 
velocity, 
\begin{equation}
v_{\infty} \ \approx \ 10,000\ R_{0.1}^{-{1\over 2}} 
\left({{{f_{0.1} L_{46}}\over{N_{22}}} - 
0.8 M_9}\right)^{{1}\over{2}} \ \ \ {\rm \kms}
\end{equation}
where 
$N_{22}$ is the total column density in $10^{22}$~\cmsq ,  
$M_9$ is the central black hole mass relative to 10$^9$~\Msun , 
and $f_{0.1}$ is the absorbed fraction compared to 10\%. 
These expressions hold strictly for 
open geometries, where the photons escaping one location in the 
wind are not scattered or absorbed in another location. 
Estimates of the total absorption indicate that $f_{0.1}$ could be as 
large as a few for BAL flows, and proportionately less for the 
narrower mini-BALs and NALs. Estimates of $N_{22}$ range 
from $\lap$0.01 for some intrinsic NALs to $\gap$1 for typical BALs. 
Radiative acceleration therefore 
requires small radii --- very roughly similar to the radius of the 
broad emission line region. 

The radial scale is important for defining 
the mass and kinetic energy of outflowing gas. 
The total wind mass at any instant is, 
\begin{displaymath}
M_w\ \approx\ 1.1\, Q_{0.1}\, N_{22}\, R_{0.1}^2 \ \ \ {\rm \Msun} 
\end{displaymath}
where $Q_{0.1}$ is the global covering factor relative to 10\%. The mass 
loss rate, $\dot M_w$, depends further on the characteristic flow 
time, $v/\Delta R$, such that,
\begin{displaymath}
\dot M_w\ \approx\ 0.11\, Q_{0.1}\, N_{22}\, 
{{R_{0.1}^2}\over{\Delta R_{0.1}}}\, v_4 \ \ \ {\rm \Msun\ yr}^{-1}
\end{displaymath}
where $v_4$ is the flow velocity in units of 10$^4$ \kms\ and 
$\Delta R_{0.1}$ is its radial thickness in units of 0.1 pc. The 
kinetic energy luminosity is $L_K \approx {{1}\over{2}}\dot M_w v^2$, or, 
\begin{displaymath}
L_{K}\ \approx\ 4\times 10^{42}\, Q_{0.1}\, N_{22}\,  
{{R_{0.1}^2}\over{\Delta R_{0.1}}}\, v_4^3 \ \ \ {\rm ergs\ s}^{-1}
\end{displaymath}
During a quasar's lifetime, perhaps $\sim$10$^8$ yrs, outflows 
with these parameter values will eject a 
total of $\sim$10$^7$ \Msun\ of gas with kinetic energy 
$\sim$$10^{58}$ ergs. BAL winds might often have total masses, 
kinetic energies, etc., that are an order of magnitude larger, 
based on our best estimates of $Q_{0.1}\approx 1$--3 and 
$N_{22}\gap 1$ for those outflows.

One unresolved issue is whether the outflowing gas is smoothly 
distributed or residing in discrete clouds. This seemingly 
simple question goes to the heart of the wind physics. It was 
long believed that the flows consist of many discrete clouds 
because, for example, a flow with $N_{22}\approx 1$ and 
$\Delta R_{0.1}\approx R_{0.1}\approx 1$ would have a mean 
density of only $n\approx 3\times 10^4$ \pcc . This density 
would lead to $U$ values several orders of magnitude larger 
than expected from the data (Eqn. 2). Moreover, 
a gas with this high ionization cannot be radiatively 
accelerated because the ions would be too stripped of electrons to 
absorb enough incident flux. The flows must 
therefore be distributed in much denser clouds that fill only 
part of the wind volume. If these clouds individually have velocity 
dispersions close to the sound/thermal speed (roughly 15 \kms\ 
for a nominal 15,000 K gas), then a smooth BAL profile of width 
$10^4$ \kms\ requires $\gap$1000 clouds along the line 
of sight. The main objection to this scenario is  
that the clouds must be very small, $\lap$10$^9$ cm across, 
and they cannot survive as discrete entities without 
some ad hoc external pressure. 

An alternative model has emerged in which the flows are, in fact, 
smoothly distributed with high ionization parameters. 
The high $U$ values, $\gap$100, 
are reconciled with the data (and with radiative acceleration) 
by invoking a large column density 
($N_{\rm H}\gap 10^{23}$ \cmsq\ for BAL winds) 
of highly ionized gas at the base of the flow. This gas is 
not radiatively accelerated because it is too ionized. 
But its bound-free absorption (in soft X-rays and the extreme UV) 
greatly ``softens'' the spectrum seen by the wind material 
downstream, thereby lowering the wind's ionization level and 
facilitating its acceleration. The most compelling 
aspect of this model is that it provides a physical basis
for the observed correlation between UV and X-ray absorption: 
the winds revealed by the UV lines cannot exist without the 
shielding X-ray absorber. The leading alternative explanation, which 
simply equates the 2 absorbing regions, presents a serious problem 
because the high column densities implied by the X-ray absorption 
might be impossible to radiatively accelerate (Eqn. 4).

A serious challenge to all models is the kinematic complexity. 
It is surprising, for example, that no intrinsic absorbers have 
shown line-of-sight acceleration. The crossing time for a flow 
with $v_4 = 2$ and $\Delta R_{0.1} = 1$ is only $\sim$5 yrs, 
yet repeated observations over this time frame reveal no velocity 
changes. Just as challenging are the multi-component 
line troughs and the frequent lack of absorption near zero velocity. 
These characteristics might be caused by the episodic ejection 
of discrete ``blobs,'' or by well-collimated flows that cross 
our line of sight and thus reveal only 
part of their full velocity extent. The latter hypothesis seems 
more likely for at least the broader lines, because their 
super-sonic velocity dispersions ($>$1000 \kms ) should 
quickly dissipate discrete blobs. Collimated accretion disk 
winds might cross our line of sight if they are driven at first 
vertically off the disk, before being bent into fully radial 
motion by the AGN's radiative force. The collimation 
and vertical ejection might both be facilitated by magnetic fields 
running perpendicular through the disk plane. However, multiple line 
troughs would require multiple collimated outflows. A major problem 
with this picture is that these intricate flow structures, 
which are tied to the accretion disk, must remain 
stable while the disk rotates. 

A more general problem is determining how much, and what aspects, 
of the diversity among intrinsic absorbers results simply from orientation 
effects. Do the various outflows identified by BALs, mini-BALs and 
intrinsic NALs coexist generally in AGNs? Orientation is clearly not 
the only factor. For example, the complete lack of high-velocity 
lines in Seyfert galaxies suggests a luminosity dependence, 
probably related to the requirements for radiative acceleration. 
Similarly, the relationships to AGN radio emissions suggest that 
there is some (unknown) physical connection 
between the disk-wind properties and the formation of radio jets. 

\subsection*{Element Abundances and Host Galaxy Evolution}

Our understanding of the element abundances near AGNs 
is based on general principles of stellar nucleosynthesis and 
galactic chemical evolution. All of the heavy elements from carbon 
on up are synthesized from primordial H and He in stars. 
The amounts of these elements are thus revealing of both 
the amount of local star formation and the evolutionary status 
of the galactic environment. The elements near AGNs 
specifically probe these properties in the centers of big galaxies. 
For distant quasars, the local abundance evolution might involve some 
of the first stars forming in collapsed structures after the Big Bang. 

Abundance measurements 
from absorption lines are, in principle, 
quite simple; one has only to apply appropriate correction 
factors for the ionization to convert the ionic column densities 
into relative abundances. For example, the abundance ratio 
of any two elements $a$ and $b$ can be written simply as, 
\begin{displaymath}
{a\over b}\ = \ 
{{N(a_i)}\over{N(b_j)}} {{F(b_j)}\over{F(a_i)}}
\end{displaymath}
where $N$ and $F$ are respectively the column densities and 
number fractions of element $a$ in ion state $i$, etc., 
in the absorbing gas. The $F$ values are generally 
adopted from photoionization calculations. 

The results from intrinsic NALs, and independently from 
the broad emission lines, indicate that quasar 
environments have roughly solar or higher heavy-element 
abundances out to redshifts $>$4. (Results based on the BALs 
are so far unreliable because of the problems 
with partial line-of-sight coverage mentioned earlier.)  
The local star formation must be both rapid and extensive 
to achieve these high abundances at early cosmic times. 
In particular, much of the gas must have already collapsed into 
stars by the time the quasars ``turned on'' or became 
observable --- just a few billion years after the Big Bang. 
These findings support prior expectations for the 
rapid, early-epoch evolution of massive galactic nuclei. 

\subsection*{Bibliography}

Studies of intrinsic AGN absorption lines began in the 1960s with 
measurements of BALs and associated NALs in quasars. 
The status of the field circa 1997 is summarized in  
numerous articles and reviews in 
the conference proceedings, Arav A, Shlosman I and Weymann R J (eds.) 
{\it Mass Ejection from AGN} (ASP Conference Series: San Francisco). 
More recent results on the BALs appear, with further references,  
in Hamann F 1998 Broad PV Absorption in the QSO 
PG 1254+047: Column Densities, Ionizations and Metal Abundances in 
Broad Absorption Line Winds {\it Astrophys. J.} {\bf 500} 798--809. 
The latest results on intrinsic X-ray absorption are discussed 
by Brandt W N, Laor A, \& Wills B J 1999 On the 
Nature of Soft X-Ray Weak Quasi-Stellar Objects 
{\it Astrophys. J.} {\bf 528} 637. The status of the polarization work 
on BAL quasars is described by Schmidt G D \& Hines D C 1999 
The Polarization of Broad Absorption Line QSOs {\it Astrophys. J.} 
{\bf 512} 125, and by Ogle P M, et al. 1999 Polarization 
of Broad Absorption Line QSOs. I. A Spectropolarimetric Atlas {\it 
Astrophys. J. Suppl.} {\bf 125} 1--34. A recent summary of 
intrinsic UV absorption in Seyfert galaxies appears in 
Crenshaw D M, Kraemer S B, Boggess A, Maran S P, 
Mushotzky R F, \& Wu C-C 1999 Intrinsic Absorption Lines in Seyfert 1 
Galaxies: I. Ultraviolet Spectra from the Hubble Space Telescope 
{\it Astrophys. J.} {\bf 516} 750--768. 
The work on AGN element abundances is 
reviewed by Hamann F H \& Ferland G J 1999 Elemental 
Abundances in QSOs: Star Formation and Galactic Nuclear Evolution at 
High Redshifts {\it Ann. Rev. Astr. Astrophys.} {\bf 37} 487.

\subsection*{Author's Credit}

Frederick Hamann \\
University of Florida

\end{document}